\documentclass[prl,twocolumn,showpacs]{revtex4}
\usepackage{graphicx}
\usepackage{amsmath}
\usepackage{epsfig}
\usepackage{psfig}
\usepackage{epsfig}

\newcounter{saveeqn}%
\newcommand{\beq}{\begin{eqnarray}}
\newcommand{\eeq}{\end{eqnarray}}
\newcommand{\beqa}{\begin{eqnarray}}
\newcommand{\eeqa}{\end{eqnarray}}

\newcommand{\bfg}{\begin{figure}}
\newcommand{\efg}{\end{figure}}

\def\pard{\partial}

\def\alp{\alpha}

\def\gam{\gamma}
\def\sig{\sigma}
\def\lam{\lambda}
\def\ome{\omega}

\def\ggg{g}

\def\kbf{{\bf k}}
\def\pbf{{\bf p}}
\def\rbf{{\bf r}}
\def\Abf{{\bf A}}
\def\Pbf{{\bf P}}
\def\Rbf{{\bf R}}

\def\omet{{\tilde \ome}}
\def\omeb{{\bar \ome}}

\begin{document}

\title{Strongly Coupled Matter-Field and Non-Analytic Decay Rate of Dipole Molecules in a Waveguide
 }
\author
{T. Petrosky$^a$$^b$,  Chu-Ong Ting$^a$, and Sterling Garmon$^a$}
\address       {         \\ $^a$Center for Studies in Statistical Mechanics and Complex Systems,
         \\ The University of Texas at Austin, Austin, TX 78712 USA 
                   \\ and
         \\ $^b$International Solvay Institutes for Physics and Chemistry,
         \\ CP231, 1050 Brussels, Belgium}

\date{\today}




\begin{abstract}
The decay rate $\gam$ of an excited dipole molecule inside a waveguide is evaluated for the strongly coupled matter-field case near a cutoff frequency $\ome_c$ without using perturbation analysis. Due to the singularity in the density of photon states at the cutoff frequency, we find that $\gam$ depends non-analytically on the coupling constant $\ggg$ as $\ggg^{4/3}$.  In contrast to the ordinary evaluation of $\gam$ which relies on the Fermi golden rule (itself based on perturbation analysis), $\gam$ has an upper bound and does not diverge at $\ome_c$ even if we assume perfect conductance in the waveguide walls. As a result, again in contrast to the statement found in the literature, the speed of emitted light from the molecule does not vanish at $\ome_c$ and  is proportional to $c\ggg^{2/3}$ which is on the order of $10^3 \sim 10^4$ m/s for typical dipole molecules.
\end{abstract}

\pacs{42.50.-p, 42.50.Pq, 42.50.Ct   \qquad \qquad \qquad \qquad \qquad \qquad \qquad                              petrosky@physics.utexas.edu
}

\maketitle


It has been well known for many years that enhanced and inhibited spontaneous emission can be observed for an atom in a resonator above and below a cutoff frequency \cite{Purcell, Kleppner}.  An explanation of this alternation has been proposed by Kleppner based on the combination of the Fermi golden rule for the transition rate and a change of the density of photon states in a resonator.  In his calculation, the singularity in the density of states at a cutoff frequency $\ome_c$ may lead to a vast increase of the decay rate $\gam$ of an excited atom or molecule compared to the free-space value. This prediction has been verified by  several experiments \cite{Drexhage}-\cite{Heinzen}. However, Kleppner's approach, while going in the right direction, is inconsistent.  This is apparent because the Fermi golden rule is not applicable in the domain of a vast increase in the decay rate, as this rule is only valid  for a weakly coupled matter-field system where the perturbation analysis may be applied.  Due to this inconsistency, $\gam$ diverges in Kleppner's theory for the case in which the resonator is made of a perfectly conducting material   and  the characteristic frequency $\ome_1$ of the atom  approaches $\ome_c$.   The decay rate $\gam$  then vanishes for $\ome_1 < \ome_c$. 

The purpose of this Letter is to re-evaluate $\gam$  for a strongly coupled case around $\ome_c$ without using perturbation analysis. The result shows that $\gam$ is finite at $\ome_c$ even if we assume  perfect conductance.  We find that the  maximum value of $\gam$   depends non-analytically on a dimensionless coupling constant $\ggg$ as $g^{4/3}$ (instead of $g^2$ which is assumed in Fermi's golden rule).    For small $g \ll 1$, $\gam$ is enhanced by a factor of $\ggg^{-2/3} $ over  the free-space value of $\gam$  that  is proportional to $g^2$.  Moreover, the frequency value at which the unstable solution disappears is found not at $\ome_1=\ome_c$, but at $\ome_1=\ome_s$, which lies slightly below $\ome_c$ with $\ome_c - \ome_s \sim g^{4/3}\ome_c$.   Again, this value is much larger than the value $\sim g^2$ predicted by the perturbation analysis for the system where the density of states has no singurality at $\ome_c$.

To present the results,
let us consider a dipole molecule (such as HCl, NaCl or KBr) that is on the order of a nanometer 
in 
size and which has a charge of $+Ze$ on one end and charge $-Ze$ on the 
other end.  This charge couples the dipole to the electromagnetic field 
in 
the rectangular waveguide.   The waveguide runs parallel to the $z$-axis and 
extends to  infinity in either diection. We refer to the width of the 
waveguide in the $x$-direction as $a$ and  to the height of the waveguide 
in 
the $y$-direction as $b$. We assume $b \ge a$.  The 
origin of $z$ is chosen at an arbitrary point along the infinite 
waveguide.   We choose the origin for $x$ and $y$ at the lower-left 
corner 
of the cross-section of the waveguide.  We assume that the center of 
mass 
of the dipole is initially located near the center of the waveguide at 
coordinates $(a/2,b/2,0)$. 
  The characteristic frequency of vibration 
$\ome_{1}$ for a typical dipole is on the order of $10^{14}$ to $10^{13}$ Hz, corresponding to a wavelength $\lam$ with
$10$ to $100\mu$m, which is in the infrared range.  Since the width of the 
waveguide is much larger than the size of the dipole, we can neglect 
the 
forces acting on the dipole from the walls of the waveguide. 

Under these conditions, one may write the Hamiltonian of the system with  the reduced mass $\mu\equiv m_am_b/(m_a+m_b)$  for the dipole and with $Z_a=-Z_b=Z$ as
\beqa
H 
& =& \sum_{i=a,b}{1 \over 2 m_i}\left( {\bf p}_i+Z_ie{{\bf A}({\bf r}_i)\over c}
\right)^2 + {1 \over 2} \mu \ome^2_1 |{\bf r_b}-{\bf r_a}|^2 \qquad  \nonumber \\
& & + {1\over 8\pi}
\int d^3 r \left( {1\over c^2} \left|{\pard {\bf A}({\bf r}) \over \pard t}
\right|^2
+|\vec{\bigtriangledown} \times {\bf A}({\bf r})|^2 \right),
\label{ori.wg.ham}
\eeqa
where ${\bf r} = (x,y,z)$. 
The vector potential inside the waveguide consists of the TE modes (Transverse Electric field modes) and TM modes (Transverse Magnetic field modes), ${\bf A} =
{\bf A}_{TE}  +{\bf A}_{TM}$ \cite{Jackson}.     For the case of the waveguide made of a perfectly conducting material, the solution of the sourceless Maxwell equations with the Coulomb gauge  inside the rectangular waveguide  leads to  these modes in terms of the normal coordinates of the fields $q_{\sigma {\bf k}} $ (with $\sigma = E$ or $M$ and $\kbf = (m, n, k)$ with $m$, $n$ integers and the continuous variable $k$) as \cite{Chu-Ong}
\beq
\mathbf{A}_{TE}(\rbf) 
&=&
 \int_{\Sigma} d\kbf \sqrt{  2I_{mn}C_{mn} \over c \ome_{\kbf}   } [-\frac{n\pi}{b}W_{1,{\kbf}}  \hat{e}_{x} \quad \nonumber \\ 
& & + \frac{m\pi}{a}W_{2,\kbf}  \hat{e}_{y}]q_{E\kbf} + h.c.,\qquad \\
\mathbf{A}_{TM} (\rbf)  
&=& \int_{\Sigma} d\kbf \frac{2c}{\pi  }
 \sqrt{  {  C_{mn} \over  \ome_\kbf^3  }    }
[i\frac{km\pi}{a}W_{1,\kbf} \hat{e}_{x}  \qquad \nonumber \\
& & + i\frac{kn\pi}{b}W_{2,\kbf} \hat{e}_{y} 
+ \alpha_{mn}^{2}W_{3,\kbf} \hat{e}_{z}]q_{M\kbf} + h.c.,  \qquad
\eeq
where $W_{1,\kbf}(\rbf) 
  \equiv \cos ({m\pi x / a} ) $ $\sin({n\pi y / b})$ $ \exp[{ikz}]$,
$W_{2,\kbf}(\rbf)$ 
$\equiv \sin ({m\pi x / a} )$ $ \cos({n\pi y / b})$ $ \exp[{ikz}]$,
$W_{3,\kbf}(\rbf)$ 
$\equiv \sin ({m\pi x / a} ) \sin({n\pi y / b})$ $ \exp[{ikz}]$, 
and $\hat{e}_i$ are unit vectors.  We put $C_{mn}\equiv 2   c^3 / (a b \alpha^2_{mn}) $ with $\alpha_{mn} \equiv  \sqrt{({m \pi / a})^2 + ({n \pi / b})^2}$,  $\ome_\kbf \equiv c \sqrt{ k^2 + \alpha^2_{mn}}$,   $\int_\Sigma d\kbf  \equiv\sum^{\infty}_{m,n \ge 0} \int^{\infty}_{-\infty} dk$, and  $I_{mn} \equiv 1$ for $m$ and $n\not=0$, $I_{mn} \equiv 1/2$ for $m$ or $n=0$.

We may rewrite the Hamiltonian in
terms of the relative coordinate ${\bf r}_1 \equiv{\bf r}_b - {\bf r}_a$, the center-of-mass coordinate ${\bf R}$, 
 and their canonical conjugate momenta $\pbf_1$ and $\Pbf$.  In this discussion we are interested in distances much larger
than the size of the molecule,  so that  the particle interacts with the field  at approximately  the center-of-mass (the so-called dipole
approximation).
 We assume that the velocity of the center-of-mass
is so slow that we can neglect its kinetic energy. 
We also assume  the field is weak enough  that we can neglect  terms that are
proportional to $\Abf^2$. 
We first consider the case in which the dipole oscillates in the $x$ direction (thus $\rbf_1
=x_1\hat{e}_x$ and  $\pbf_1
=p_1\hat{e}_x$).  The extension to arbitrary direction will be  discussed later.  We
then introduce the unperturbed normal coordinate $q_1\equiv \sqrt{\mu \ome_1 / 2}
[x_1+i(p_1/ \mu \ome_1)]$ of the dipole. 
With these assumptions, the
Hamiltonian (\ref{ori.wg.ham}) may be approximatelly written in a bilinear form with respect to the annihilation and creation operators (which are related to the normal coordinate through $q_\alp \equiv \sqrt{\hbar}a_\alp$) as
\beqa
H &=&
  \hbar\ome_1 a^+_1 a_1 +\sum_{\sig}^{
 E,M}
\int_\Sigma d\kbf \hbar\ome_{\kbf} a^+_{\sig {\bf k}} a_{\sig  {\bf
k}} \qquad \nonumber \\
& & +\ggg\sum_{\sig}^{E,M}
\int_\Sigma d\kbf  ( V_{\sig,{\bf k}} 
a_{\sig {\bf k}}-V_{\sig,{\bf k}}^* 
a^+_{\sig {\bf k}})
(a_1-a_1^+),
\label{ham.fried}
\eeqa
with  the dimensionless coupling constant $\ggg\equiv \sqrt{ (Ze)^2 \ome_1 / (\mu_1 c^3)}$.   We have $\ggg \sim 10^{-6}$ to  $10^{-7}$ for the typical dipole molecules.
The interactions are given by
$V_{E,{\bf k}}
\equiv
-i({n/ b}) \sqrt{I_{mn} }
F_{1,\kbf}(\Rbf) , 
$
and
$V_{M,{\bf k}} 
\equiv
 - ({m c  k /a \ome_\kbf }) F_{1,\kbf}(\Rbf)
$, 
with $F_{1,\kbf}(\Rbf) \equiv \hbar \pi^2 C_{mn}$ $W_{1,\kbf}(\Rbf)/\sqrt{\ome_\kbf}$. The operators $a_\alp$ satisfy  the usual   commutation relations.
For each $(m,n)$ mode, the continuous spectrum $\ome_\kbf$  is
bounded from below at $c \alpha_{mn}$. These lower bounds 
 form a set of cutoff frequencies  in the sense that only electromagnetic
modes with frequency  $\ome_\kbf > c \alpha_{mn} $ may propagate inside
the waveguide.
Among these branches of continuum, the TE mode with $m=0$ and $n=1$ has the
smallest value for its cutoff frequency $\ome_c \equiv c\alp_{01} = \pi c/b$.

The Hamiltonian  (\ref{ham.fried}) has exactly the same structure as the Hamiltonian for the well-known Friedrichs model with virtual processes \cite{Karpov, NewFried}.   This model has been investigated extensively in order to analyze the spontaneous decay of an excited atom \cite{Karpov}-\cite{TPRad}.  Since the Hamiltonian is bilinear, one can find its exact diagonal form  as 
$H 
 = \hbar \omeb_1 b_1^+ b_1 + \sum_{\sig }\int_\Sigma d\kbf\, \hbar\ome_{\bf k} b^+_{\sig  {\bf k}} b_{\sig {\bf k}},
$
where $\omeb_1$ is the shifted real frequency for the stable dressed harmonic oscillator.   The new dressed annihilation operators $b_\alp$ are obtained by  the Bogoliubov transformation  and satisfy the usual commutation relations.    The explicit form of the Bogoliubov transformation is presented in \cite{Chu-Ong,Karpov}. 

In this short letter we limit our studies to  the shifted frequency and the decay rate of the harmonic oscillator. These are found by solving the dispersion equation $\xi (z ) =0$ for the dressed harmonic oscillator, where 
\beq
\xi (z) 
\equiv z^2-\ome^2_1 -  \ggg^2
\sum_{\sig} \int_\Sigma d\kbf 
 \frac{4\ome_1\ome_{\kbf} { |V_{\sig,{\bf k}} |}^2}{z^2-\ome^2_k}.
\label{wg.def.xi}
\eeq
The solution of $\xi (z) =0$ gives a real solution $z = z_0 \equiv  \omeb_1$  with a shifted  frequency  for the stable mode of the harmonic oscillator  and  a complex solution $z= z_1 \equiv  \omet_1 - i \gam$  with  shifted frequency $\omet_1$ and  decay rate $\gam$ for the unstable mode. 

 To find the explicit form of these solutions, let us consider the case in which $\ome_1$ is located  below the next smallest cutoff frequency $\ome_{c2}$.   The geometry of the waveguide can be  chosen in such a way that the cutoff frequency $\ome_c$ is well separated from $\ome_{c2}$, and as such, we may have  $\ome_1 \le \ome_c$, or  $\ome_c < \ome_1< \ome_{c2}$ where $\ome_1$ is much closer to  $\ome_c$  than $\ome_{c2}$.  For this case the predominant contribution to the integration in Eq. (\ref{wg.def.xi}) comes from the component with $\sig=E$, $m=0$ and $n=1$.  Approximating the integration by leaving out all other components and explicitly performing  the integration,  we obtain  the dispersion equation \cite{integ} 
\beq  \label{Disp}
\zeta - w_1^2 
 =
\mp\frac{  \ggg^{2}G^2w_1 }{ \sqrt{1 - \zeta } } ,
\eeq
where we have introduced the  dimensionless variables, $w_1 \equiv \ome_1/ \ome_c$, $\zeta\equiv (z/\ome_c)^2$, and a dimensionless constant $
G^2  \equiv  (4c/a\ome_c) \sin^2(\pi R_y/b)
$.
The minus branch of Eq. (\ref {Disp}) gives the real solution $z=z_0$ for the stable mode, while the plus branch gives  the complex solution $z=z_1$ for the unstable mode.   In the following discussion we shall consider the case where the dipole remains near  the center of the waveguide, without loss of generality, and put
$G_0 \equiv G|_{R_y=b/2}$. Squaring  Eq. (\ref{Disp}), we obtain a cubic equation for $\zeta$.  Using the standard method to solve the cubic equation, one can explicitly find the shifted frequency and the decay rate.  However, since  the equation was squared, we must take care  to exclude non-physical solutions.  The discriminant of the cubic equation is given by
$D(w_1)  =  4p_{w_1}^3 +q_{w_1}^2$,
where 
\beq \label{pDefn}
p_w
\equiv   - \frac{1}{3^2}(w^2 - 1)^{2},
\quad
q_w
\equiv \frac{2}{3^3}(w^2 - 1)^3 + \ggg^{4}G_0^4w^2.
\eeq
The stability of the dipole is determined by the critical frequency at $w_1 =w_s$, which is given by $D(w_s)=0$.
This gives
\beq
w_s^2
   & =& 1 - \frac{ 3\ggg^{8/3}G_0^{ 8/3} }{ 2( \sqrt{1 + \ggg^{4}G_0^4} -1)^{1/3} } \qquad \nonumber \\
& & + \frac{3}{2}\ggg^{4/3}G_0^{4/3}( \sqrt{1 + \ggg^4G_0^4 } -1)^{1/3}.
\eeq
For $\ggg\ll 1$ we have $\ome_{s} \approx \ome_c
[1 - 2^{1/3} (3/4)({\ggg G_0 })^{4/3} ] $ where $ \ome_s \equiv w_s \ome_c$.
For a given value of $ \ome_c$ only the stable  mode  exists with real $z=z_0$ for  $\ome_1 < \ome_s$, while   both the stable  mode  and the unstable mode with a complex $z =z_1$ exist for $\ome_1 > \ome_s$. 

Then,  we define $
\alp_{w\pm}
 \equiv (-q_w \pm \sqrt{q_w^{2}+4p_w^{3}})/2
$ in order to write the solutions of Eq.(\ref{Disp})  as     
\beq \label{Soln}
z_n = \ome_c \sqrt{  e^{{2ni\pi/ 3}}\alp_{w_1+}^{1/3} + e^{-{2ni\pi/ 3}}\alp_{w_1-}^{1/3} + \frac{1}{3}(2w_{1}^{2} + 1) },
\eeq
where $n=0$ for the stable solution and $n=1$ for the unstable solution.

\begin{figure}[htb] 
\begin{center}
\includegraphics[width=3.5in]{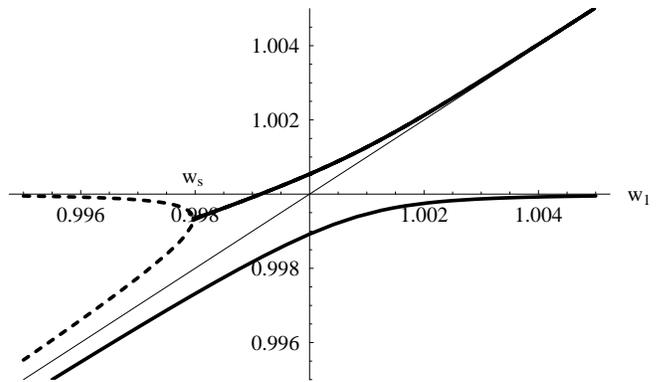}
\caption{ The vertical axis represents the shifted frequencies ${\bar w}_1 \equiv \omeb_1/\ome_c$   and ${\tilde w}_1 \equiv \omet_1/\ome_c$, and the transverse axis is  $w_1\equiv \ome_1/\ome_c$ with the value of $\ome_c$ fixed.      We indicate the location of the critical value $w_s $.  The thin line is $y=w_1$. The thick curve below the thin line  is ${\bar w}_1$ for the stable mode.  The thick curve above the thin line  is ${\tilde w}_1$  for the unstable mode. The two dashed  curves  are unphysical  real solutions which correspond to $n=1$ and $n=2$ in Eq. (\ref{Soln}) for $w_1< w_s$.} 
\label{Fig1} \end{center}
\end{figure}

In Fig. 1 we plot the value of ${\bar w}_1 =\omeb_1/\ome_c$   and ${\tilde w}_1 =\omet_1/\ome_c = {\rm Re}(z_1/\ome_c)$  as functions of $w_1=\ome_1/\ome_c$ with a fixed value of $\ome_c$.     In this  and the next  figures  we set  $G_0 =2$ and use a large  coupling constant $\ggg=0.005$  to exaggerate the effect of the interaction.     We indicate the location of the critical value $w_s $.  The thin line is $y=w_1$. The thick curve below the thin line  is ${\bar w}_1$ for the stable mode.  The thick curve above the thin line  is ${\tilde w}_1$  for the unstable mode. The two dashed  curves in the domain  $w_1< w_s$ are unphysical  real solutions obtained from $z_n$ by putting $n=1$ and 2.

We note that the stable solution $\bar{w}_1$ exists for all values of $w_1$ inside the waveguide.   This is the result of the singularity in the density of  states at the cutoff frequency $\ome_k =\ome_c$ in Eq. (\ref{wg.def.xi}) appearing as $cdk = \ome_kd\ome_k/(\ome_k^2-\ome_c^2)^{1/2}$.   Because of this singularity, there may be  a large deviation  in the value of $\bar{w}_1$ from $w_1$ for $w_1 \gg 1$ no matter how small the coupling constant $g$ might be.  This is a striking difference from the ordinary Friedrichs model which is used to analyze the spontaneous emission of the photon from the atom located in the vacuum without boundary \cite{Pronko}.  In contrast to $\bar{w}_1$, the deviation  of $\tilde{w}_1$ from $w_1$ for the unstable mode is always small for $\ggg\ll1$. 

 In Fig. 2 we plot the decay rate $\gam/\ome_c ={ \rm Im} \, (z_1/\ome_c) $ of the unstable mode for a fixed value of $\ome_c$ as a function of $w_1$.   The maximum value of $\gam$ in our system is obtaind at $w_1=1$ (i.e., $\ome_1=\ome_c$).  At this point we have $p_{w_1}=\alp_{w_1+} =0$, $q_{w_1}=\alp_{w_1-} =(\ggg G_0)^4$, and we obtain for $\ggg \ll 1$,
\beq \label{gammax}
\gam_{max}  
= \frac{\sqrt{3}}{4}\ggg^{4/3}\Big( {4b\over \pi a}\Big)^{2/3}\ome_c  + O(\ggg^{8/3}).
\eeq
The maximum value is a nonanalytic function at $\ggg=0$, and hence one cannot obtain this result from perturbation analysis.   For  $\ggg \ll 1$ this is much learger than  the decay rate $ \gam_2 \equiv  2\ggg^2 (c/a)(\ome_c/\ome_1)$  which is found in the perturbation region where
 $|w_1 -1|/ |w_s -1 | \gg 1$.   Indeed, the enhanced factor for the decay rate $\gam_{max} / \gam_2 \propto\ggg^{-2/3}$  is  extremely large  for  the case  $w_1 >1$ with $w_1 \sim 1$ and $\ggg\ll 1$.

Notice that the critical value $w_s$ of the unstable mode is located below the cutoff frequency, i.e., $w_s < 1$.   This is another striking difference from the ordinary Friedrichs model, in which the critical value is located inside the continuous spectrum of the field \cite{Pronko}.  

\begin{figure}
\begin{center}
\includegraphics[width=3.5in]{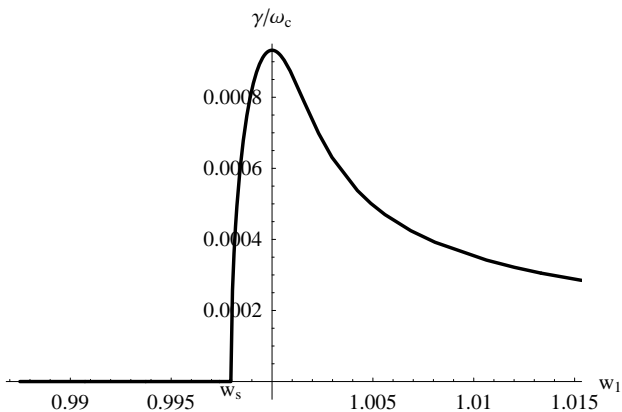}
\caption{Decay rate  $\gam/\ome_c$ of the unstable mode as a function of $w_1=\ome_1/\ome_c$ for a fixed value of $\ome_c$.  The decay rate does not diverge at $w_1 =1$  even though we have assumed the waveguide is made of a perfectly conducting material.   At $w_1 =1$ we have the maximum value of $\gam$, which  depends non-analytically on the coupling constant $g$ as presented at Eq. (\ref{gammax}). } 
\label{Fig2} \end{center}
\end{figure}

We   note that $\gam$ does not depend of the size of the molecule in the dipole approximation.
One can show that the vibrating motion of the dipole is stable when  it oscillates in the $y$ direction, or in the $z$ direction for  our illustrated case with $\ome_1 < \ome_{c2}$.  As a result,  we have the same value of $\gam$ for molecules  oriented in any direction of space.

We also note that the $\ggg^{4/3}$ law obtained in our system is rather universal around the cutoff frequency, as it can be shown to be independent of the particular model we choose.
 Indeed, one can find the $\ggg^{4/3}$ law by estimating the lower bound of $\ome_1$ at which the Fermi golden rule is applicable in the vicinity of the singularity in the density of states at $\ome_c$. Due to lack of space in this Letter, we will present this estimation elsewhere.
Finally, we remark on the  velocity $v_f$ of the light emitted from the dipole. Through the exact form of the Bogoliubov transformation, we find
\beqa \label{vg}
v_f =
 {c \gam 
       \over
\sqrt{
\sqrt{   {1\over 4} (\omet_1^2 - \gam^2 -\ome_c^2)^2 + \omet_1^2\gam^2 }
 -{1\over 2}(\omet_1^2 - \gam^2 -\ome_c^2) 
}. 
} \,\,\,
\eeqa
In the perturbation region $|w_1 -1|/ |w_s -1 | \gg 1$ in which  $|\omet_1 - \ome_1| \sim \ggg^2 \ome_1$ and $\gam \sim \ggg^2 \ome_1$, this expression reduces to the well-known formula for the ordinary group velocity $v_f \approx v_g \equiv c[1 - (\ome_c/\ome_1)^2]^{1/2}$ inside the waveguide \cite{Jackson}.    However,  since the perturbation analysis  fails at $\ome_1=\ome_c$, one should not conclude from the form of $v_g$ that the group velocity vanishes at the cutoff frequency as stated, for example, in \cite{Jackson}.    Indeed, Eq. (\ref{vg}) leads to  $v_f \approx 2cg^{-2/3}\gam_{max}  \approx (\sqrt{3}/2)g^{2/3} c$ at $\ome_1=\ome_c$ for the group velocity.  For typical dipole molecules, $v_f$ at $\ome_1=\ome_c$ is on the order of $10^3 \sim 10^4$ m/s, which is comparable with the speed of phonons propagating on the walls of the waveguide.

 We thank Professor H. Jeff Kimble for his helpful comments regarding the relation of our result to the experiment presented in  \cite{Hultet}.
We acknowledge  the
Engineering Research Program of the Office of Basic Energy Sciences at 
the U.S. Department of Energy, Grant No DE-FG03-94ER14465 and the European Commission Grant No
HPHA-CT-2001-40002 for supporting  this work.


\vfil\eject


\begin{thebibliography}{99}
\bibitem{NewFried} The original model discussed by Friedrichs in \cite{Fried} was the case of a point spectrum coupled with a continuum that corresponded with a fermionic particle coupled with a field.  The Hamiltonian considered here is an extention of  the Friedrichs model for a bosonic partilce.
\bibitem{integ}  The integration  over $k$ in (\ref{wg.def.xi}) leads to a factor $(\ome_{mn}^2 - z^2)^{-1/2}$ for each TE mode, while  $ [(\ome_{mn}^2 - z^2)^{-1/2} - \ome_{mn}^{-1}]/z^2$ for each TM mode, where $\ome_{mn} \equiv c\alp_{mn}$ is the cutoff frequency for each mode.
\end{thebibliography}
 \end{document}